\documentclass[usenatbib,useAMS,onecolumn]{mn2e}\usepackage[]{graphicx}\usepackage[]{color}
\makeatletter
\def\maxwidth{ %
  \ifdim\Gin@nat@width>\linewidth
    \linewidth
  \else
    \Gin@nat@width
  \fi
}
\makeatother

\definecolor{fgcolor}{rgb}{0.345, 0.345, 0.345}

\usepackage{framed}
\makeatletter
 {\par\unskip\endMakeFramed%
 \at@end@of@kframe}
\makeatother

\definecolor{shadecolor}{rgb}{.97, .97, .97}
\definecolor{messagecolor}{rgb}{0, 0, 0}
\definecolor{warningcolor}{rgb}{1, 0, 1}
\definecolor{errorcolor}{rgb}{1, 0, 0}
\newenvironment{knitrout}{}{} 

\usepackage{alltt}

\usepackage{aas_macros}
\usepackage{amsmath}
\usepackage{amssymb}
\usepackage{graphicx}
\usepackage{longtable}
\usepackage{mathptmx}
\usepackage[section]{placeins}

\pagerange{\pageref{firstpage}--\pageref{lastpage}} \pubyear{2014}

\IfFileExists{upquote.sty}{\usepackage{upquote}}{}

\begin{document}

\title[Flicker noise pulsar radio spectra]{Flicker noise pulsar radio spectra}

\author[Krzeszowski et. al.]{K.~Krzeszowski,$^1$ O.~Maron,$^1$ A.~S\l{}owikowska$^1$
and A.~Jessner$^2$ \\
$^1$Kepler Institute of Astronomy, University of Zielona G\'ora, Lubuska 2, 65--265,
Zielona G\'ora, Poland\\
$^2$Max-Planck-Institut f\"ur Radioastronomie, Auf dem H\"ugel 69, D-53121 Bonn, Germany\\
}
\date{Released 2014}
\maketitle

\begin{abstract}
We present new results of fitting 108 spectra of 
radio pulsars with the flicker noise model
proposed by \citet{loehmer08} and compare them with the spectral indices of power--law 
fits published by \citet{maron2000}. The fits to the model were carried out using
the Markov chain Monte Carlo (MCMC) method appropriate for the non--linear fits.
Our main conclusion is that pulsar radio spectra can be statistically very well
described by the flicker noise model over wide frequency range from a few tens of MHz
up to tens of GHz. Moreover, our dataset allows us to conduct statistical analysis
of the model parameters. As our results show, there is a strong negative correlation
between the flicker noise spectrum model parameters $\log S_0$ and $n$ and a strong positive
relationship between $n$ and the power--law spectral index $\alpha$. The latter implies that their physical meaning
is similar, however the flicker noise model has an advantage over broken power--law
model. Not only it describes the spectra in higher frequency range with only two
parameters, not counting scaling factor $S_0$, but also it shows smooth transition
from flat to steep behaviour at lower and higher frequencies, respectively.
On the other hand, there are no correlations of the flicker
noise model parameters $S_0$, $\tau$ and $n$ with any of pulsar physical properties.
\end{abstract}

\label{firstpage}


\section{Introduction}
\label{sec:introduction}
Spectra of radio pulsars have been a subject of studies for many years. It is a generally
accepted fact, that the pulsar radio spectra are steep and approximately follow
a power-law relation: $S_{\nu} \sim \nu^{\alpha}$, where $S_{\nu}$ is the
observed flux density at frequency $\nu$, and $\alpha$ is the spectral index.

Many studies have been carried out on the measurements of flux densities in
a wide frequency range over the past years \citep[e.g.][]{sieber73, malofeev94,
seiradakis95, lorimer95, maron2000}. However, the first systematic study of
spectral indices for a large number of pulsars was conducted by \citet{lorimer95}
who published a catalogue of spectral indices for 280 pulsars based on observations 
made with the Lovell radio telescope at frequencies 0.4--1.6~GHz. 
More recent work by \citet{maron2000} extended the frequency range
for many cases up to 5~GHz and derived the spectral index value of
$\alpha=-1.8 \pm 0.2$. They also found that in their sample $\sim 10\%$ of
pulsars have double power law (broken power law) type spectrum which
is described by two parameters, $\alpha_1$ at lower frequencies and
$\alpha_2$ at higher frequencies. Additionally, \citet{maron2000} discovered
that PSRs B1823-13 and B1838-04 show high-frequency turnovers in their spectra.
While turnovers are often observed at frequencies around and below 100 MHz
\citep{izvekova81}, the occurrence of such an effect at frequency
${\rm \sim 1 GHz}$ is unusual.

In this paper we are investigating the flicker noise model
proposed by \citet{loehmer08} where the received radio flux density is given by
\begin{equation}
{\rm S(\omega)=S_0\left(\frac{1+\omega^2\tau^2}{\tau^2}\right)^{n-1}\!\!\!\!\times e^{-i(n-1)atan(\omega\tau)},}
\label{equation:spectrum}
\end{equation}
where $S_0$ is a scaling factor, ${\rm \omega=2\pi\nu}$, $\nu$ is an observing
frequency, $\tau$ is the characteristic life time of nano bursts and $n$ is the exponent that captures a number 
of physical and observational properties of the emission \citep{loehmer08}. Pulsar fluxes and hence the input 
values for pulsar spectra are commonly averaged over the pulse period and therefore represent an average over what we 
receive  from different emission regions at different phase angles. We know that many sources have strong profile 
evolution which is hidden in the single flux values and it would be rather surprising if a single one parameter 
law can fit such a spectrum well. Nevertheless, L\"ohmer et al. have pointed out, that even a two parameter model 
for a reference radio flux and a single time scale can already provide surprisingly good fits to the spectrum with 
intuitive values for the timescales that could be interpreted as observed life times of nano pulses. If one was to 
assume similar conditions for the emission regions observed at different pulse phases, as one would perhaps imagine 
to exist in pulsars with simple one component profiles, then  $n$ could have more than a heuristic meaning. Assume that 
we have  radiation processes that are confined to flux tubes forming a layer of depth $\delta$, having a particular 
cross section $\sigma$ and typical separation $a$ and that these emission centres can interact destructively if they 
overlap, then $n={\sigma \over \pi a \delta}$. If the local plasma wavelength 
$\lambda_{pe}(r,\gamma)= {2\pi \over \omega_{pe}(r,\gamma)}$  is the decisive scale, then $n$ would just be thickness 
of the emission layer or alternatively the separation of emission centres in units of plasma lengths. Our results show, 
that like $\tau$ the range of $n$ is constrained which points to similar conditions for the emission processes of all 
pulsars even when one takes the averaging over phase angles and propagation effects into account.

Section~\ref{sec:fitting} covers the data selection and fitting method description.
In Section~\ref{sec:results} we discuss our findings which we summarise and
conclude in Section~\ref{sec:summary}. 

\section{Dataset and fitting method}
\label{sec:fitting}
Our base dataset for the analysis consists of flux measurements of 108 pulsars. The sample was selected
from a much wider dataset, containing measurements for 281 pulsars, published by \citet{maron2000}. 
The most important data selection criterion was the coverage
of the possible widest range of  frequencies --- flux measurements should be available at low ($<$200~MHz) and high ($>$1~GHz) frequencies to
unambiguously fit all the parts of L\"ohmer spectrum model. Only those pulsars that met this criterion were included in
the sample. Single datasets consist of minimum of 4 up to maximum of 24 
flux measurements at a number of frequencies with the mean of 8 measurements.
In some of the cases the fits were not reasonable due to insufficient number of low and/or high frequency measurements
or measurements that were doubtful i.e. outliers or unusually high errors. Such cases are not presented in this paper, and after collection of additional 
published flux measurements will be presented in the upcoming paper.

\subsection{MCMC}
\label{sec:mcmc}
The model fits according to Eq.~\ref{equation:spectrum} were carried out using the Markov chain Monte Carlo (MCMC) method which 
is a machine learning algorithm \citep{gelfand90}. It is mostly used to solve 
optimisation problems in multidimensional spaces, although it can be used for solving much simpler problems. It is a 
combination of memoryless Markov process and Monte Carlo randomised algorithm. The Markov process fulfils the condition, 
that future state of the system depends only on its present state without a need for knowing its full history. One
common example is the random walk where next position depends only on present position regardless of previous steps.
The main idea behind MCMC is to construct a Markov chain in the way, that its equilibrium distribution is the distribution 
we want to sample. To achieve this goal a number of so called walkers randomly wander the parameter space and draw
a set of samples from the original, maybe unknown, distribution. After high number of steps we sample the chain 
instead of sampling the original distribution. 

In case of a non--linear model like a power--law, one can not use $\chi^2_\mathrm{red}$ as the goodness of fit estimator 
since determination of degrees of freedom is non trivial \citep{andrae10}. As the fitting procedure we used a maximum 
likelihood method along with Markov chain Monte Carlo using \textit{emcee} python library \citep{mcmc}. 

The fitting procedure was run on datasets consisting of measured values of observing frequency, flux and flux
error. The MCMC algorithm in connection with the maximum likelihood estimation (MLE) allowed us to sample 
the $\mathbb{R}^3$ parameter space sufficiently to get smooth parameter distributions. We took the median 
(50\textsuperscript{th} percentile) of the resulting distributions of each parameter as parameter values as well as 
16\textsuperscript{th} and 84\textsuperscript{th} percentiles of the distributions as a lower ($\sigma^-$) and upper 
($\sigma^+$) uncertainties, respectively. For the normal distribution the 16th and 84th percentiles 
are equivalent of 2 sigma range and they are commonly used for error estimation.

\section{Results and discussion}
\label{sec:results}

As an example of our analysis the fitted spectrum to PSR~B1133+16 flux measurements is presented in Fig.~\ref{fig:spectrum}.
The same approach was
taken in case of 108 sets of flux measurements for different pulsars and yielded estimations of parameters 
$S_0$, $\tau$ and $n$. Parameter distributions for PSR~B1133+16, along with their correlations, 
as an example are presented in Fig.~\ref{fig:triangle}.
The plot presents 1--D and 2--D posterior probability distributions of three fitted parameters $S_0$, $\tau$ and $n$. The
most frequently occurring values for each parameter are marked with solid vertical and horizontal lines. The diagonal
1--D distributions are symmetric and Gaussian--like which is not always like this. In this case it means, that the
fit is reasonable and the parameter estimations along with their errors, taken as the 16\textsuperscript{th} and
84\textsuperscript{th} percentile of each distribution, are precise.

Resulting parameters $S_0$, $\tau$ and $n$ for all pulsars and respective uncertainties, along with a number of flux measurements
of fitting 108 spectra are gathered in~Tab.~\ref{tab:results}. The table also contains spectral indices from \citet{maron2000}
of power--law fits (95 values) and high frequency parts of broken power--law fits (13 values, denoted
with superscript \textit{b}). The distributions of 
all fitted parameters and their upper and lower uncertainties are shown in~Fig.~\ref{fig:histograms}. The parameters range from 0.015~Jy
(B2022+50) to 14.127~Jy (B0531+21)
in case of $S_0$, from 0.041~ns (B1800-21) to 2.014~ns
(B2022+50) in case of $\tau$ and from 0.003 (B2000+40) to
0.544 (B2022+50) in case of $n$ with mean values of 1.197~Jy, 0.586~ns
and 0.143 for $S_0$, $\tau$ and $n$, respectively. Fifty percent of $S_0$, $\tau$ and $n$ values are between
0.17--1.20~Jy, 0.25--0.84~ns and 0.07--0.20, respectively. One of the most noticeable pulsar in our sample is PSR~B2022+50 (J2023+5037). 
It has a period of P = 0.373~s., a polarisation angle curve showing an
orthogonal mode and its interpulse is 100\% polarised \citep{han09}. Our fit of the flicker noise spectrum model to B2022+50
flux measurements yielded $S_0$ = 0.015~Jy, $\tau$ = 2.014~ns
and $n$ = 0.544, which are extreme values within our whole sample.
This pulsar is the weakest one so the scaling factor $S_0$ of its fit is also the smallest among all values. It is supposed to
have the longest duration of nano--bursts ($\tau$) and its spectrum is not as steep as of other pulsars due to the high $n$ value. 

\citet{loehmer08} fitted flicker noise spectra to 12 pulsars and got $S_0$, $\tau$ and $n$. In Tab.~\ref{tab:comparison}
we compare them with our results. Our results are consistent with previous studies except
for B0144+59 and B1929+10 where $\tau$ is different by the order of magnitude. Since there are no plots or data for
these pulsars in their work, we can not investigate either of the cases in details.

We also looked for the correlations between fitted parameters $S_0$, $\tau$ and $n$ and found a relationship between $S_0$ and
$n$ with Pearson's correlation index of -0.52 with the 95\% confidence interval of (-0.65, -0.37) and the p--value of \ensuremath{5.53\times 10^{-9}}.
Remaining parameters show only weak
correlations, as it can be seen in Fig.~\ref{fig:new.cor}. In the diagonal of this figure the approximated kernel densities of parameter
distributions are plotted. The main pulsar radio spectrum model fitted in \citet{maron2000} was a power--law with the mean value of spectral index
$\alpha = -1.8 \pm 0.2$. Correlations of $S_0$, $\tau$ and $n$ acquired in this study with spectral indices $\alpha$ from fits of
single power--law are presented in Fig.~\ref{fig:correlation.plots}. There are no obvious correlations between $S_0$ and $\tau$ with $\alpha$.
However, there is a strong positive relationship between $n$ and $\alpha$ with Pearson's correlation index of 0.63
with the 95\% confidence interval of (0.50, 0.73) and the p--value of \ensuremath{2.03\times 10^{-13}}. The linear fit yields the relation 
$n$~=~0.11~$\alpha$~+~0.34. This suggests that the physical interpretation of $n$ and the 
power--law spectral index $\alpha$ is similar. Apart from that, there are no correlations of $S_0$, $\tau$ and $n$ with any of pulsar physical 
 parameters like period, period derivative, age, rotational energy loss, magnetic field on surface or dispersion and rotation measures.

\subsection{PSR~B1133+16}
\label{sec:1133}

PSR~B1133+16 is one of the strongest and most frequently observed pulsars. Since 2000 there were many observations which allowed
us to extend the original data (19 measurements) from \citet{maron2000} with additional 41 measurements that were 
gathered in \citet{krzeszowski14}. Fig.~\ref{fig:spectrum} shows the results of fitting the flicker noise spectrum to PSR~B1133+16 flux measurements. 
These data were divided into two sets: original set of points from \citet{maron2000} marked with black filled circles and all measurements
from \citet{krzeszowski14} marked with black filled circles and black opened circles altogether. The fits are plotted with solid line
and dashed line for Maron's data and Krzeszowski's data, respectively, however both fits are indistinguishable. 
The spectrum is relatively flat at lower frequencies ($<$ 300~MHz) and becomes steep at higher frequencies.
The resulting parameters are as follows: 
$S_0^\mathrm{M} = 3.48^{+0.34}_{-0.35}$~Jy, 
$\tau^\mathrm{M} = 0.39^{+0.03}_{-0.03}$~ns, 
$n^\mathrm{M} = 0.12^{+0.01}_{-0.01}$
and  
$S_0^\mathrm{K} = 3.39^{+0.35}_{-0.32}$~Jy, 
$\tau^\mathrm{K} = 0.40^{+0.02}_{-0.02}$~ns, 
$n^\mathrm{K} = 0.12^{+0.01}_{-0.01}$, where superscripts M and K denote respective
datasets. In this case it is clear that tripling the number of available data points from 19 to 60
does not alter the fit when the original data sample the frequency space sufficiently well in low and high frequency regions.

In \cite{krzeszowski14} we fitted L\"ohmer model to 60 flux measurements using the least squares method
and got the following results: $S_0 = 3.39 \pm 0.77~\rm Jy$,
$\tau = 0.40 \pm 0.05~\rm  ns$, $n = 0.118 \pm 0.022$ and $\chi^2_\mathrm{red} = 4.7$.
The model is non--linear and the obtained $\chi^2_\mathrm{red}$
is much greater than 1 and the error distribution is not normal, hence least squares method is not appropriate 
here (see Sec.~\ref{sec:mcmc}). Still, the values
are in good agreement with the values obtained with the MCMC, however
the parameter errors from MCMC are 2--2.5 times smaller than in case of the least squares method.

\begin{longtable}{lrrrrrrrrrrl}
\caption{The fitted parameters ($S_0$ [Jy], $\tau$ [ns] and $n$) with their respective upper ($\sigma^+$) and lower ($\sigma^-$) 
 uncertainties, N denotes a number of flux measurements that fit was performed on. Additionally, power--law spectral index $\alpha$
 is included for reference \citep{maron2000}. $\alpha_2$ values from broken power--law are denoted with superscript \textit{b}.} \\ 
  \hline
Pulsar & N & $S_0$ & $\sigma_{S_0}^+$ & $\sigma_{S_0}^-$ & $\tau$ & $\sigma_{\tau}^+$ & $\sigma_{\tau}^-$ & n & $\sigma_{n}^+$ & $\sigma_{n}^-$ & $\alpha$ \\ 
  \hline 
\endhead 
\hline 
{\footnotesize Continued on next page} 
\endfoot 
\endlastfoot 
 \hline
B0011+47 &    6 & 0.200 & 0.162 & 0.083 & 0.219 & 0.097 & 0.049 & 0.214 & 0.086 & 0.092 & -1.3 \\ 
  B0037+56 &    7 & 0.056 & 0.007 & 0.006 & 1.096 & 0.302 & 0.224 & 0.193 & 0.017 & 0.017 & -1.8 \\ 
  B0052+51 &    7 & 0.170 & 0.073 & 0.047 & 0.124 & 0.019 & 0.015 & 0.077 & 0.042 & 0.043 & -0.7 \\ 
  B0105+65 &    4 & 0.086 & 0.032 & 0.023 & 0.641 & 0.210 & 0.144 & 0.154 & 0.074 & 0.075 & -1.9 \\ 
  B0136+57 &   11 & 0.457 & 0.049 & 0.045 & 0.219 & 0.019 & 0.016 & 0.126 & 0.013 & 0.013 & -2.3$^b$ \\ 
  B0138+59 &    9 & 0.573 & 0.069 & 0.056 & 0.432 & 0.052 & 0.053 & 0.115 & 0.015 & 0.016 & -1.9 \\ 
  B0144+59 &   10 & 0.024 & 0.004 & 0.004 & 1.901 & 0.718 & 0.660 & 0.415 & 0.031 & 0.031 & -1.0 \\ 
  B0148-06 &    5 & 0.196 & 0.089 & 0.069 & 0.933 & 0.883 & 0.382 & 0.169 & 0.066 & 0.066 & -2.7 \\ 
  B0301+19 &    9 & 0.416 & 0.211 & 0.123 & 0.298 & 0.083 & 0.063 & 0.106 & 0.062 & 0.061 & -2.3$^b$ \\ 
  B0320+39 &    8 & 0.384 & 0.047 & 0.049 & 0.887 & 0.133 & 0.136 & 0.048 & 0.009 & 0.008 & -2.9 \\ 
  B0329+54 &   24 & 10.085 & 0.827 & 0.852 & 0.263 & 0.014 & 0.014 & 0.077 & 0.006 & 0.006 & -2.2 \\ 
  B0331+45 &    6 & 0.054 & 0.019 & 0.013 & 0.917 & 0.716 & 0.350 & 0.260 & 0.055 & 0.061 & -1.9 \\ 
  B0353+52 &    7 & 0.156 & 0.028 & 0.023 & 0.309 & 0.055 & 0.045 & 0.148 & 0.023 & 0.024 & -1.6 \\ 
  B0355+54 &   21 & 0.689 & 0.455 & 0.246 & 0.085 & 0.041 & 0.020 & 0.398 & 0.041 & 0.052 & -1.2$^b$ \\ 
  B0402+61 &    5 & 0.132 & 0.019 & 0.013 & 0.457 & 0.297 & 0.095 & 0.233 & 0.019 & 0.022 & -1.4 \\ 
  B0410+69 &    5 & 0.078 & 0.011 & 0.010 & 1.220 & 0.337 & 0.245 & 0.097 & 0.020 & 0.019 & -2.4 \\ 
  B0450+55 &   10 & 0.366 & 0.047 & 0.037 & 0.558 & 0.227 & 0.186 & 0.272 & 0.012 & 0.014 & -1.5 \\ 
  B0458+46 &    7 & 0.241 & 0.033 & 0.027 & 0.240 & 0.018 & 0.016 & 0.094 & 0.019 & 0.020 & -1.3 \\ 
  B0523+11 &    8 & 0.200 & 0.018 & 0.016 & 0.772 & 0.128 & 0.128 & 0.144 & 0.017 & 0.019 & -2.0 \\ 
  B0531+21 &    6 & 14.127 & 3.408 & 2.734 & 1.631 & 0.620 & 0.438 & 0.019 & 0.012 & 0.010 & -3.1 \\ 
  B0540+23 &   11 & 0.808 & 0.265 & 0.170 & 0.119 & 0.015 & 0.014 & 0.232 & 0.026 & 0.029 & -1.6$^b$ \\ 
  B0626+24 &    6 & 0.255 & 0.035 & 0.030 & 1.145 & 0.533 & 0.330 & 0.206 & 0.024 & 0.024 & -1.6 \\ 
  B0628-28 &   13 & 2.696 & 0.223 & 0.206 & 0.550 & 0.047 & 0.047 & 0.103 & 0.009 & 0.009 & -1.9 \\ 
  B0643+80 &    6 & 0.062 & 0.014 & 0.011 & 0.526 & 0.143 & 0.116 & 0.157 & 0.029 & 0.031 & -1.9 \\ 
  B0740-28 &   11 & 2.991 & 0.216 & 0.193 & 0.527 & 0.042 & 0.039 & 0.110 & 0.011 & 0.012 & -2.0 \\ 
  B0809+74 &   13 & 0.998 & 0.097 & 0.099 & 1.322 & 0.160 & 0.143 & 0.211 & 0.010 & 0.010 & -1.4 \\ 
  B0823+26 &   16 & 0.577 & 0.073 & 0.070 & 0.946 & 0.138 & 0.126 & 0.211 & 0.012 & 0.012 & -2.1$^b$ \\ 
  B0834+06 &   11 & 1.767 & 0.141 & 0.162 & 1.032 & 0.078 & 0.072 & 0.011 & 0.007 & 0.006 & -2.7 \\ 
  B0919+06 &   14 & 0.499 & 0.067 & 0.065 & 0.814 & 0.131 & 0.111 & 0.162 & 0.018 & 0.018 & -1.8 \\ 
  B0950+08 &   14 & 12.619 & 0.787 & 0.842 & 0.304 & 0.022 & 0.021 & 0.061 & 0.004 & 0.004 & -2.2 \\ 
  B1112+50 &    7 & 0.064 & 0.010 & 0.009 & 1.752 & 0.778 & 0.575 & 0.293 & 0.026 & 0.026 & -1.7 \\ 
  B1133+16 &   19 & 3.482 & 0.342 & 0.345 & 0.393 & 0.030 & 0.028 & 0.115 & 0.008 & 0.008 & -1.9 \\ 
  B1237+25 &   15 & 2.980 & 0.182 & 0.170 & 0.256 & 0.018 & 0.017 & 0.056 & 0.006 & 0.006 & -2.2$^b$ \\ 
  B1508+55 &   11 & 1.431 & 0.115 & 0.103 & 0.676 & 0.054 & 0.053 & 0.054 & 0.015 & 0.015 & -2.2 \\ 
  B1540-06 &    7 & 0.213 & 0.033 & 0.031 & 0.934 & 0.273 & 0.189 & 0.208 & 0.026 & 0.027 & -2.0 \\ 
  B1541+09 &    8 & 1.222 & 0.071 & 0.064 & 0.540 & 0.106 & 0.120 & 0.023 & 0.006 & 0.009 & -2.6 \\ 
  B1552-23 &    5 & 0.101 & 0.031 & 0.021 & 0.322 & 0.083 & 0.054 & 0.142 & 0.041 & 0.046 & -1.8 \\ 
  B1604-00 &   11 & 0.467 & 0.091 & 0.078 & 0.626 & 0.102 & 0.089 & 0.143 & 0.024 & 0.025 & -1.5 \\ 
  B1612+07 &    6 & 0.137 & 0.028 & 0.022 & 1.214 & 0.451 & 0.393 & 0.078 & 0.010 & 0.010 & -2.6 \\ 
  B1633+24 &    5 & 0.101 & 0.013 & 0.013 & 0.851 & 0.190 & 0.165 & 0.103 & 0.022 & 0.023 & -2.4 \\ 
  B1642-03 &   12 & 3.468 & 0.500 & 0.495 & 0.423 & 0.035 & 0.031 & 0.044 & 0.015 & 0.015 & -2.3 \\ 
  B1702-19 &    7 & 0.145 & 0.025 & 0.022 & 1.952 & 0.682 & 0.635 & 0.333 & 0.027 & 0.026 & -1.3 \\ 
  B1706-16 &   12 & 0.284 & 0.031 & 0.030 & 0.842 & 0.191 & 0.185 & 0.275 & 0.010 & 0.010 & -1.5 \\ 
  B1730-22 &    4 & 0.432 & 0.069 & 0.061 & 0.358 & 0.113 & 0.062 & 0.043 & 0.018 & 0.018 & -2.0 \\ 
  B1732-07 &    4 & 0.369 & 0.044 & 0.040 & 0.311 & 0.036 & 0.030 & 0.051 & 0.015 & 0.016 & -1.9 \\ 
  B1737+13 &    6 & 0.209 & 0.043 & 0.032 & 0.587 & 0.110 & 0.101 & 0.199 & 0.033 & 0.037 & -1.5 \\ 
  B1738-08 &    6 & 0.394 & 0.073 & 0.069 & 0.703 & 0.186 & 0.153 & 0.089 & 0.019 & 0.018 & -2.2 \\ 
  B1742-30 &    6 & 0.641 & 0.089 & 0.078 & 0.363 & 0.076 & 0.051 & 0.247 & 0.021 & 0.021 & -1.6 \\ 
  B1749-28 &   10 & 8.925 & 0.536 & 0.742 & 0.449 & 0.026 & 0.022 & 0.005 & 0.009 & 0.004 & -4.3$^b$ \\ 
  B1750-24 &    5 & 0.095 & 0.075 & 0.037 & 0.096 & 0.020 & 0.015 & 0.261 & 0.069 & 0.078 & -1.0 \\ 
  B1753-24 &    5 & 0.350 & 0.049 & 0.067 & 0.121 & 0.012 & 0.009 & 0.025 & 0.028 & 0.017 & -0.7 \\ 
  B1800-21 &    7 & 3.449 & 3.494 & 1.369 & 0.041 & 0.006 & 0.005 & 0.178 & 0.055 & 0.070 & -1.0$^b$ \\ 
  B1804-08 &    5 & 0.958 & 0.408 & 0.225 & 0.244 & 0.040 & 0.032 & 0.198 & 0.051 & 0.062 & -1.2 \\ 
  B1811+40 &    5 & 0.104 & 0.048 & 0.029 & 0.569 & 0.233 & 0.183 & 0.165 & 0.061 & 0.070 & -1.8 \\ 
  B1815-14 &    6 & 0.897 & 0.359 & 0.274 & 0.105 & 0.016 & 0.012 & 0.070 & 0.048 & 0.043 & -1.6 \\ 
  B1819-22 &    6 & 0.561 & 0.054 & 0.047 & 0.245 & 0.020 & 0.019 & 0.036 & 0.014 & 0.015 & -1.7 \\ 
  B1820-11 &    6 & 0.152 & 0.039 & 0.029 & 0.247 & 0.049 & 0.036 & 0.289 & 0.033 & 0.035 & -1.5 \\ 
  B1822+00 &    4 & 0.119 & 0.025 & 0.025 & 0.839 & 0.342 & 0.245 & 0.060 & 0.046 & 0.038 & -2.4 \\ 
  B1826-17 &    7 & 1.913 & 0.167 & 0.155 & 0.234 & 0.017 & 0.015 & 0.045 & 0.010 & 0.010 & -1.7 \\ 
  B1830-08 &    5 & 0.446 & 0.441 & 0.180 & 0.088 & 0.016 & 0.015 & 0.188 & 0.067 & 0.085 & -1.1 \\ 
  B1831-04 &    6 & 0.465 & 0.091 & 0.061 & 0.465 & 0.110 & 0.090 & 0.311 & 0.032 & 0.038 & -1.3 \\ 
  B1839+56 &    7 & 0.254 & 0.076 & 0.054 & 0.289 & 0.097 & 0.064 & 0.188 & 0.032 & 0.036 & -1.5 \\ 
  B1841-04 &    6 & 0.157 & 0.052 & 0.045 & 0.267 & 0.059 & 0.034 & 0.083 & 0.061 & 0.050 & -1.6 \\ 
  B1844-04 &    6 & 0.762 & 0.053 & 0.046 & 0.620 & 0.058 & 0.055 & 0.091 & 0.014 & 0.015 & -2.2 \\ 
  B1846-06 &    5 & 0.323 & 0.031 & 0.027 & 0.749 & 0.151 & 0.127 & 0.080 & 0.018 & 0.020 & -2.2 \\ 
  B1857-26 &    6 & 2.346 & 0.223 & 0.208 & 0.334 & 0.041 & 0.032 & 0.035 & 0.014 & 0.015 & -2.1 \\ 
  B1859+03 &    6 & 3.134 & 0.510 & 0.376 & 1.229 & 0.255 & 0.204 & 0.013 & 0.006 & 0.006 & -2.8 \\ 
  B1859+07 &    6 & 0.312 & 0.081 & 0.091 & 0.134 & 0.020 & 0.012 & 0.041 & 0.040 & 0.027 & -1.0 \\ 
  B1900-06 &    4 & 0.317 & 0.057 & 0.044 & 0.528 & 0.303 & 0.116 & 0.067 & 0.040 & 0.042 & -1.8 \\ 
  B1900+05 &    5 & 0.179 & 0.039 & 0.025 & 0.439 & 0.331 & 0.102 & 0.174 & 0.029 & 0.036 & -1.7 \\ 
  B1900+06 &    6 & 0.352 & 0.061 & 0.043 & 0.402 & 0.212 & 0.083 & 0.058 & 0.023 & 0.026 & -2.2 \\ 
  B1903+07 &    4 & 0.165 & 0.043 & 0.032 & 0.185 & 0.029 & 0.024 & 0.137 & 0.032 & 0.033 & -1.3 \\ 
  B1904+06 &    4 & 0.141 & 0.178 & 0.061 & 0.131 & 0.060 & 0.044 & 0.246 & 0.075 & 0.103 & -0.7 \\ 
  B1907+02 &    5 & 0.285 & 0.030 & 0.027 & 0.992 & 0.152 & 0.136 & 0.040 & 0.014 & 0.015 & -2.8 \\ 
  B1907+03 &    5 & 0.215 & 0.056 & 0.047 & 1.448 & 0.821 & 0.476 & 0.190 & 0.030 & 0.028 & -1.8 \\ 
  B1907+10 &    5 & 0.778 & 0.062 & 0.059 & 0.768 & 0.119 & 0.102 & 0.018 & 0.012 & 0.011 & -2.5 \\ 
  B1911-04 &    5 & 1.249 & 0.105 & 0.098 & 0.497 & 0.039 & 0.037 & 0.060 & 0.012 & 0.012 & -2.6 \\ 
  B1913+10 &    8 & 0.283 & 0.034 & 0.031 & 1.195 & 0.315 & 0.256 & 0.129 & 0.012 & 0.011 & -1.9 \\ 
  B1915+13 &    8 & 0.533 & 0.215 & 0.143 & 0.380 & 0.184 & 0.082 & 0.120 & 0.064 & 0.063 & -1.8 \\ 
  B1916+14 &    6 & 0.023 & 0.013 & 0.008 & 0.204 & 0.077 & 0.046 & 0.317 & 0.057 & 0.062 & -0.3 \\ 
  B1919+21 &   10 & 1.190 & 0.065 & 0.067 & 0.490 & 0.043 & 0.042 & 0.011 & 0.005 & 0.005 & -2.6 \\ 
  B1923+04 &    4 & 0.219 & 0.031 & 0.027 & 0.600 & 0.136 & 0.123 & 0.108 & 0.015 & 0.016 & -2.7 \\ 
  B1929+10 &   19 & 5.355 & 0.830 & 0.895 & 0.078 & 0.022 & 0.013 & 0.163 & 0.015 & 0.013 & -1.6 \\ 
  B1933+16 &    6 & 2.666 & 0.394 & 0.367 & 0.346 & 0.082 & 0.051 & 0.192 & 0.019 & 0.018 & -1.7 \\ 
  B1935+25 &    6 & 0.112 & 0.053 & 0.032 & 0.153 & 0.027 & 0.021 & 0.282 & 0.048 & 0.054 & -0.7 \\ 
  B1946+35 &    7 & 2.454 & 0.128 & 0.122 & 0.357 & 0.015 & 0.014 & 0.021 & 0.010 & 0.010 & -2.4 \\ 
  B2000+40 &    8 & 1.009 & 0.030 & 0.043 & 0.235 & 0.015 & 0.009 & 0.003 & 0.008 & 0.002 & -2.2 \\ 
  B2011+38 &    8 & 0.541 & 0.132 & 0.094 & 0.176 & 0.017 & 0.015 & 0.162 & 0.030 & 0.032 & -1.9$^b$ \\ 
  B2020+28 &   12 & 2.511 & 1.159 & 0.724 & 0.116 & 0.021 & 0.016 & 0.159 & 0.031 & 0.032 & -1.9$^b$ \\ 
  B2021+51 &   20 & 1.661 & 0.517 & 0.519 & 0.124 & 0.048 & 0.026 & 0.269 & 0.028 & 0.022 & -1.5$^b$ \\ 
  B2022+50 &    8 & 0.015 & 0.003 & 0.003 & 2.014 & 0.675 & 0.751 & 0.544 & 0.034 & 0.034 & -0.8 \\ 
  B2044+15 &    5 & 0.076 & 0.010 & 0.008 & 1.118 & 0.536 & 0.290 & 0.236 & 0.017 & 0.018 & -1.7 \\ 
  B2045-16 &   11 & 2.235 & 0.341 & 0.347 & 0.238 & 0.030 & 0.028 & 0.069 & 0.013 & 0.012 & -2.1 \\ 
  B2045+56 &    5 & 0.064 & 0.013 & 0.012 & 0.986 & 0.313 & 0.198 & 0.111 & 0.035 & 0.031 & -2.4 \\ 
  B2053+36 &    6 & 0.265 & 0.018 & 0.015 & 0.566 & 0.070 & 0.061 & 0.161 & 0.012 & 0.013 & -1.9 \\ 
  B2106+44 &    5 & 0.328 & 0.119 & 0.073 & 0.315 & 0.071 & 0.050 & 0.181 & 0.049 & 0.056 & -1.4 \\ 
  B2110+27 &    6 & 0.257 & 0.030 & 0.027 & 0.675 & 0.170 & 0.203 & 0.079 & 0.012 & 0.013 & -2.2 \\ 
  B2111+46 &   10 & 2.996 & 0.263 & 0.218 & 0.414 & 0.036 & 0.033 & 0.076 & 0.013 & 0.014 & -2.1 \\ 
  B2148+52 &    7 & 0.080 & 0.007 & 0.006 & 1.074 & 0.697 & 0.381 & 0.333 & 0.013 & 0.014 & -1.3 \\ 
  B2148+63 &    7 & 0.465 & 0.048 & 0.040 & 0.320 & 0.021 & 0.019 & 0.079 & 0.015 & 0.016 & -1.8 \\ 
  B2154+40 &    7 & 1.392 & 0.517 & 0.303 & 0.292 & 0.074 & 0.053 & 0.128 & 0.040 & 0.047 & -1.6 \\ 
  B2217+47 &    7 & 1.389 & 0.154 & 0.151 & 0.856 & 0.086 & 0.084 & 0.058 & 0.017 & 0.017 & -2.6 \\ 
  B2227+61 &    5 & 0.178 & 0.018 & 0.016 & 0.703 & 0.065 & 0.060 & 0.072 & 0.020 & 0.021 & -2.6 \\ 
  B2303+30 &    7 & 0.155 & 0.039 & 0.033 & 1.701 & 0.698 & 0.468 & 0.195 & 0.033 & 0.030 & -2.3 \\ 
  B2310+42 &   10 & 1.072 & 0.099 & 0.087 & 0.495 & 0.063 & 0.055 & 0.093 & 0.011 & 0.012 & -1.9 \\ 
  B2319+60 &   11 & 1.459 & 0.191 & 0.163 & 0.152 & 0.013 & 0.012 & 0.125 & 0.011 & 0.012 & -2.1$^b$ \\ 
  B2324+60 &    7 & 0.618 & 0.153 & 0.177 & 0.168 & 0.026 & 0.018 & 0.044 & 0.044 & 0.029 & -2.5$^b$ \\ 
  B2351+61 &    9 & 0.168 & 0.039 & 0.028 & 0.372 & 0.083 & 0.074 & 0.220 & 0.023 & 0.026 & -1.1 \\ 
   \hline
\hline
\label{tab:results}
\end{longtable}

\begin{knitrout}
\definecolor{shadecolor}{rgb}{0.969, 0.969, 0.969}\color{fgcolor}\begin{figure}[]

\includegraphics[width=\maxwidth]{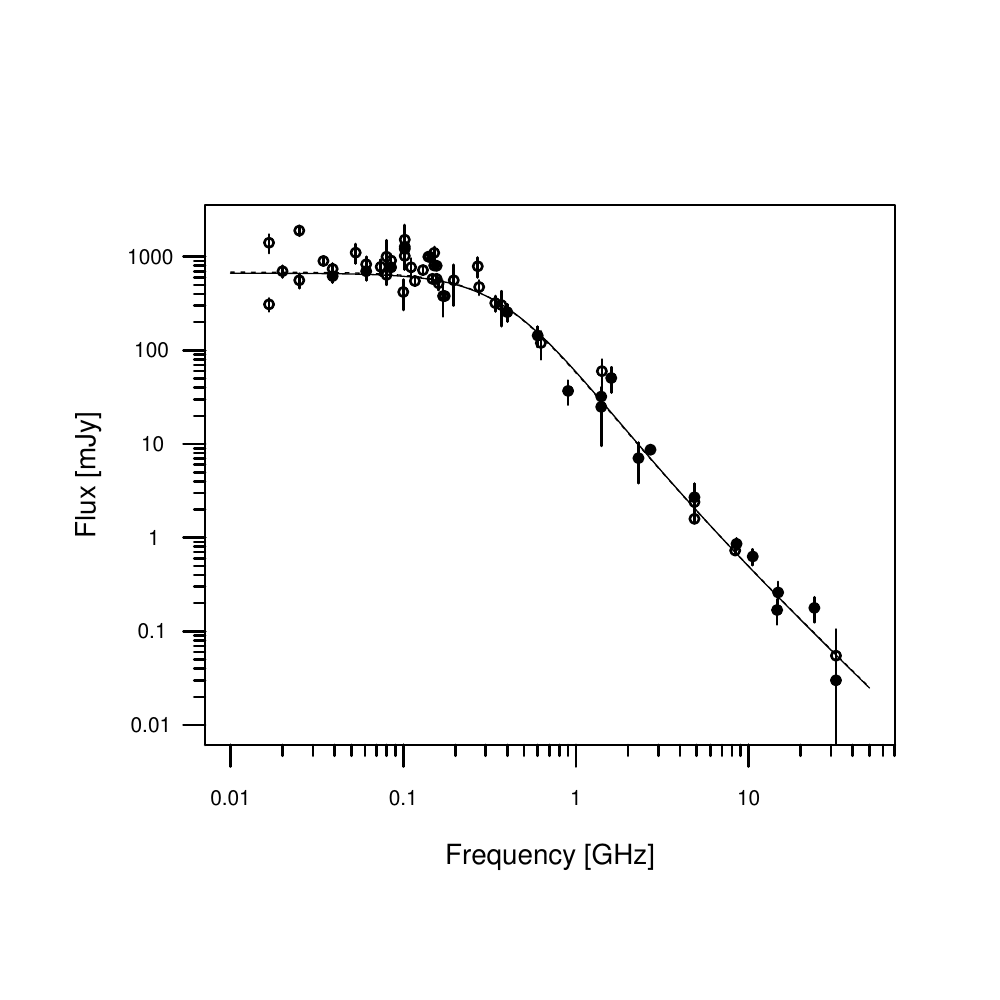} \caption[Fitted spectrum of PSR~B1133+16]{Fitted spectrum of PSR~B1133+16. Black filled circles denote 19 flux measurement from \citet{maron2000}, while black open circles denote 41 additional points gathered in \citet{krzeszowski14}. The fits are plotted with solid line and dashed line for Maron's data and Krzeszowski's datasets, respectively. Extending the original dataset of 19 with additional 41 measurements hasn't changed the fit (see sec.~\ref{sec:1133}).\label{fig:spectrum}}
\end{figure}

\end{knitrout}

\begin{table}[ht]
\centering
\caption{Comparison of the fitted values with the values in \citet{loehmer08} denoted with superscript $L$ for 12 pulsars.} 
\label{tab:comparison}
\begin{tabular}{lrrrrrr}
  \hline
Pulsar & $S_0$ & $\tau$ & $n$ & $S_0^L$ & $\tau^L$ & $n^L$ \\ 
  \hline
B0144+59 & 0.02 & 1.90 & 0.42 & 0.03 & 0.34 & 0.45 \\ 
  B0329+54 & 10.08 & 0.26 & 0.08 & 22.00 & 0.16 & 0.07 \\ 
  B0355+54 & 0.69 & 0.08 & 0.40 & 0.27 & 0.27 & 0.50 \\ 
  B0628-28 & 2.70 & 0.55 & 0.10 & 3.09 & 0.62 & 0.11 \\ 
  B0823+26 & 0.58 & 0.95 & 0.21 & 1.77 & 0.71 & 0.17 \\ 
  B0950+08 & 12.62 & 0.30 & 0.06 & 7.68 & 0.42 & 0.12 \\ 
  B1133+16 & 3.48 & 0.39 & 0.12 & 4.41 & 0.51 & 0.10 \\ 
  B1706-16 & 0.28 & 0.84 & 0.28 & 0.40 & 0.91 & 0.27 \\ 
  B1929+10 & 5.36 & 0.08 & 0.16 & 2.59 & 0.57 & 0.23 \\ 
  B2020+28 & 2.51 & 0.12 & 0.16 & 1.61 & 0.12 & 0.28 \\ 
  B2021+51 & 1.66 & 0.12 & 0.27 & 1.54 & 0.24 & 0.20 \\ 
  B2022+50 & 0.02 & 2.01 & 0.54 & 0.03 & 1.77 & 0.48 \\ 
   \hline
\end{tabular}
\end{table}

\begin{figure}
\includegraphics[width=0.5\textwidth]{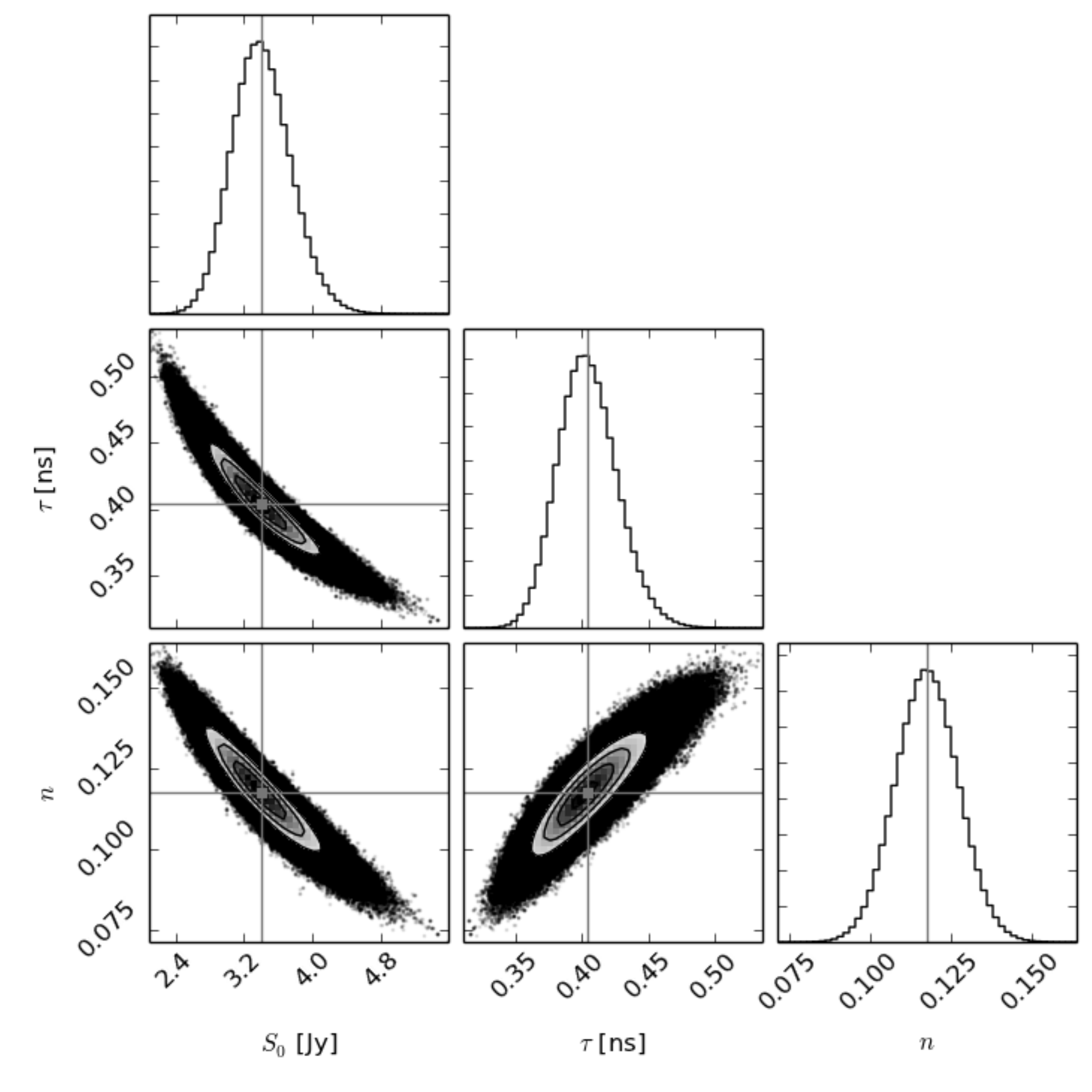}
\caption{Triangle plot of distributions of fitted parameters (diagonal) and 2--d correlations between parameters for
a number of MCMC walkers in parameter space of PSR~B1133+16 extended data (60 measurements, see Fig.~\ref{fig:spectrum}) spectrum fit. 
Solid vertical and horizontal lines denote most frequently occurring values of each parameter. The 16\textsuperscript{th} and 
84\textsuperscript{th} percentile of each distribution are lower and upper uncertainties of each parameter estimation, 
respectively.
\label{fig:triangle}}
\end{figure}

\begin{knitrout}
\definecolor{shadecolor}{rgb}{0.969, 0.969, 0.969}\color{fgcolor}\begin{figure}[]

\includegraphics[width=\maxwidth]{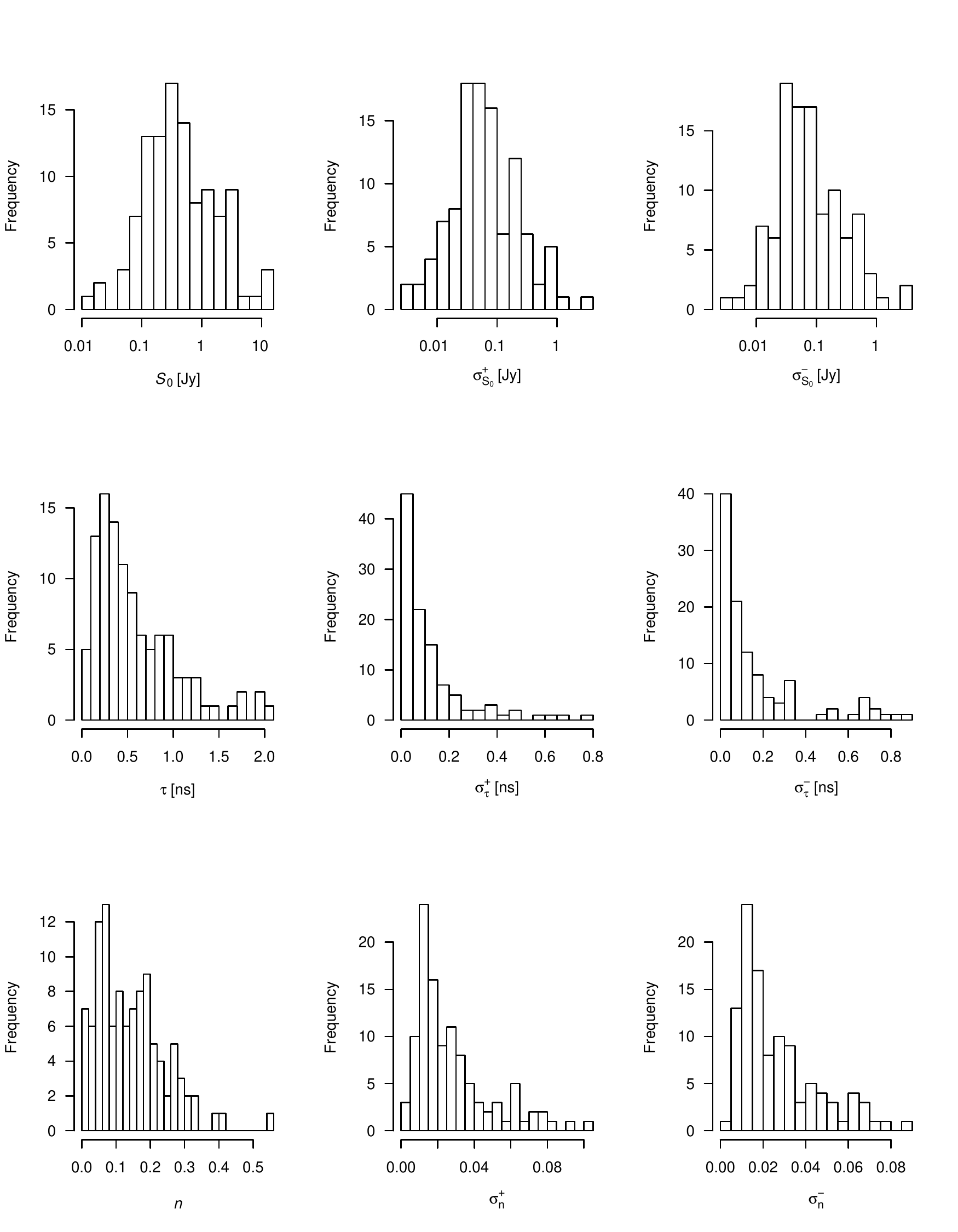} \caption[$S_0$, $\tau$ and $n$ and their upper and lower uncertainties distributions of spectra fitted to 108 pulsars flux measurements]{$S_0$, $\tau$ and $n$ and their upper and lower uncertainties distributions of spectra fitted to 108 pulsars flux measurements. Fifty percent of $S_0$, $\tau$ and $n$ values are between 0.17--1.20 ~Jy, 0.25--0.84 ~ns and 0.07--0.20 , respectively.\label{fig:histograms}}
\end{figure}

\end{knitrout}

\begin{knitrout}
\definecolor{shadecolor}{rgb}{0.969, 0.969, 0.969}\color{fgcolor}\begin{figure}[]

\includegraphics[width=\maxwidth]{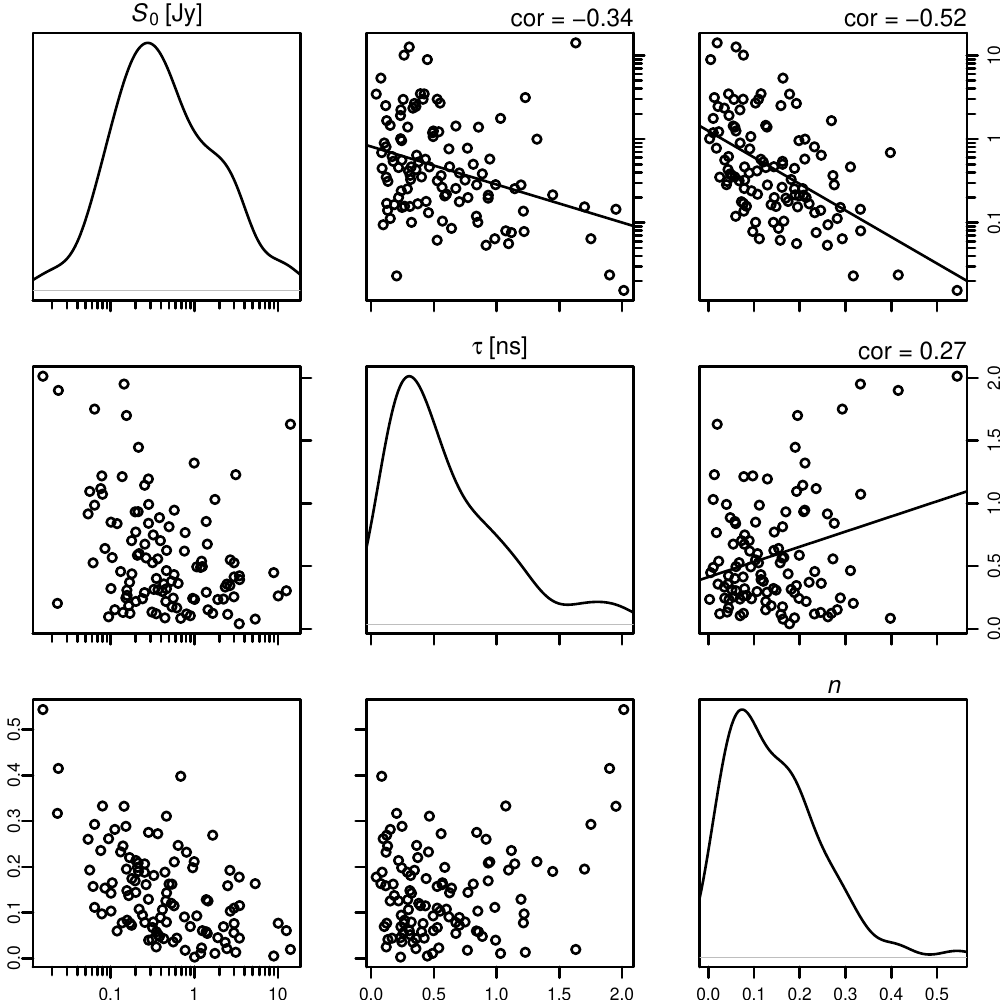} \caption[Density plots and correlations of fitted parameters $S_0$, $\tau$ and $n$]{Density plots and correlations of fitted parameters $S_0$, $\tau$ and $n$. The first row shows the radio flux density distribution, the scatter plot and correlations between $\tau$ and $S_0$ as well as $n$ and $S_0$. The second row shows the scatter plot of $S_0$ versus $\tau$, the distribution of $\tau$ and the scatter plot and correlation between $n$ and $\tau$. The third row shows the scatter plots of $S_0$ versus $n$ ans $\tau$ versus $n$ as well as the distribution of $n$. Pearson's correlation indices are printed over upper corner panels. There is a strong negative correlation between $\log S_0$ and $n$.\label{fig:new.cor}}
\end{figure}

\end{knitrout}

\begin{knitrout}
\definecolor{shadecolor}{rgb}{0.969, 0.969, 0.969}\color{fgcolor}\begin{figure}[]

\includegraphics[width=\maxwidth]{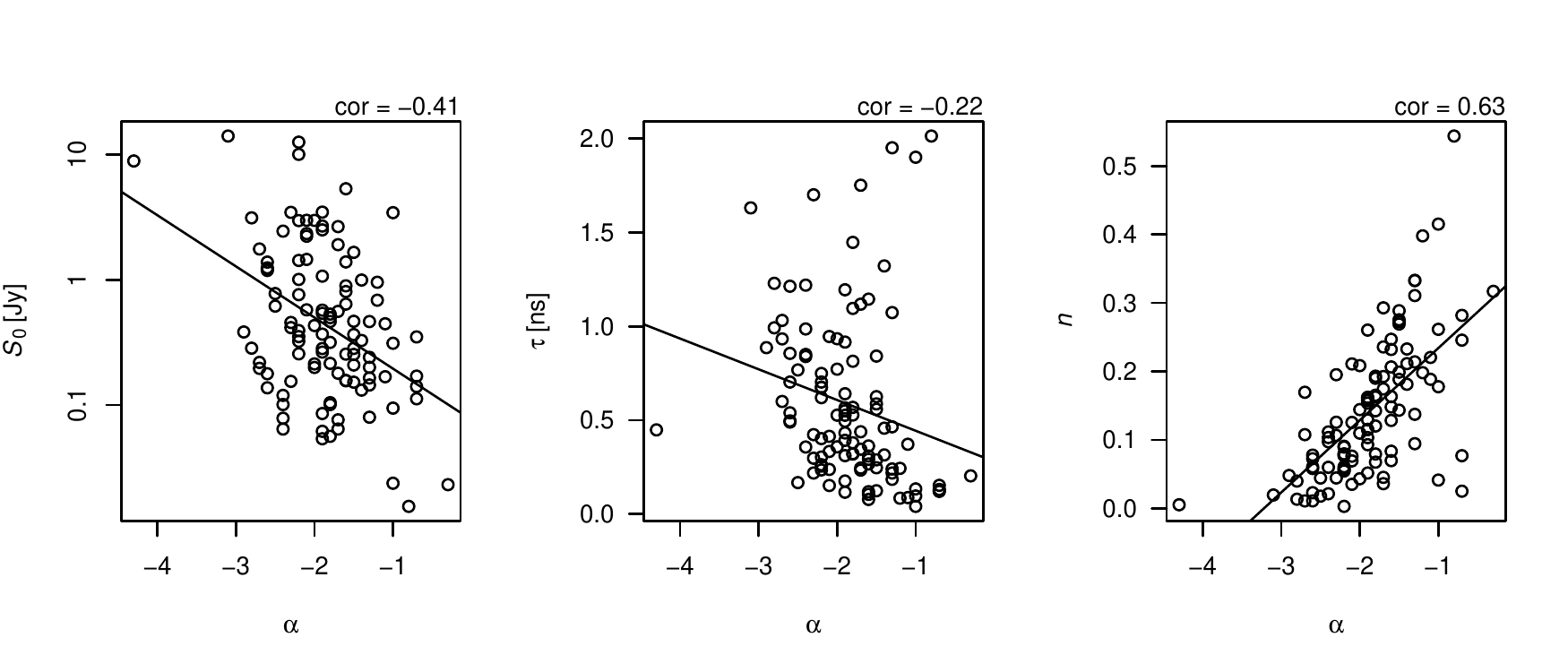} \caption[Correlation plots of the parameters $S_0$, $\tau$ and $n$ obtained from our fit with the spectral indices $\alpha$ of single power--law fits from \citet{maron2000}]{Correlation plots of the parameters $S_0$, $\tau$ and $n$ obtained from our fit with the spectral indices $\alpha$ of single power--law fits from \citet{maron2000}. There are no obvious correlations between $S_0$ and $\tau$ with $\alpha$. However, there is a strong positive relationship between $n$ and $\alpha$ with Pearson's correlation index of 0.63.\label{fig:correlation.plots}}
\end{figure}

\end{knitrout}

\section{Summary and conclusions}
\label{sec:summary}
The main conclusion of our work is that the pulsar radio spectrum can be well described with flicker noise model proposed
by \citet{loehmer08} for a high number of pulsars. The crucial constraint on the quality of fitting the flicker noise 
spectrum model is the coverage of frequency space. There should be flux measurements at low ($<$ ~200~MHz) and 
high ($>$~1~GHz) frequencies to obtain reliable  parameter estimates. We have flux measurements for many more pulsars
but they lack either low or high frequency data. We are planning to extend Maron's database with
published measurements and new observations. In the best case it will almost double our sample up to 281 
pulsars for our future study.

\citet{maron2000} fitted spectrum for flux measurements above 300~MHz even if they had data points at lower frequencies
because of so called low--frequency turn--over, which can not be reasonably represented with power--law model. In our
work we took into account all the points, including those below 300~MHz gathered by \citet{maron2000} and successfully
fitted the flicker noise spectrum model. The most outstanding example of extension of the data is the case of PSR~B1133+16,
which justifies application of L\"ohmer's model, that describes pulsar radio spectrum in wide frequency range
from 17~MHz up to 32~GHz. To reproduce more complex shape of a pulsar spectrum with power--laws one should
combine three or more of them to get a reasonable fit. This approach is artificial and most likely non-physical. 
It seems that flicker noise model has an advantage over a power--law model and also over broken power--law. Not only does it describe the spectra in higher 
frequency range with only two parameters, not counting scaling factor $S_0$, but it also shows a smooth transition from 
flat to steep behaviour at lower and higher frequencies, respectively. Based on our dataset of 108 pulsars 
we found that fifty percent of $S_0$, $\tau$ and $n$ values are between
0.17--1.20~Jy, 0.25--0.84~ns and 0.07--0.20, respectively.

We conclude that there is a strong negative correlation between the flicker noise spectrum model parameters $\log S_0$ 
and $n$, which may indicate a selection bias as high frequency detections would be improbably for weak steep spectrum 
sources, and a strong positive relationship between $n$ and the power--law spectral index $\alpha$.
The latter implies, that the physical meaning of both parameters must be similar. 
The flicker noise spectrum behaves in the similar manner as the power--law at higher 
frequencies and is able to describe the data at lower frequencies at the same time, which
is another advantage of this model. 

Furthermore, in our study we have not found any correlations of the model parameters $S_0$, $\tau$ and $n$
with pulsar physical properties like period, period derivative, age, rotational energy loss, 
magnetic field on surface or dispersion and rotation measures. 

\section*{Acknowledgements}
This work has been supported by Polish
National Science Centre grants DEC-2011/03/D/ST9/00656 (KK, AS) and DEC-2012/05/B/ST9/03924 (OM). Data analysis and
figures were partly prepared using R \citep{rcite}. We want to thank prof. Krzysztof Go\'zdziewski from CA UMK (Toru\'n,
Poland) for very useful and fruitful discussion on the MCMC application to fitting non--linear models to data.

\bibliographystyle{mn2e}
\bibliography{krzeszowski}

\bsp
\label{lastpage}
\end{document}